\documentstyle[12pt]{article}
\def\lsim{\stackrel{<}{\sim}}
\def\fax{ F_{i}(x)}

\def\ka{ \kappa}
\def\wa{ w_\alpha }
\def\difu{\sqrt{\langle\Delta S^2\rangle}}
\def\VEV#1{ \langle {#1}\rangle}
\def\partder#1#2{\frac{\partial #1}{\partial #2}}
\def\co{{\cal O}}

\begin{document}
\vspace*{1mm}
{\raggedleft YCTP-N1-92\\
             OHSTPY-HEP-T-92-013\\
             DOE/ER/01545-583\\}
\mbox{}
\vspace{1cm}
\begin{center}
{\Large\sc Global Demons in Field Theory :}\\
\vspace{3mm}
{\Large\sc Critical Slowing Down in the XY Model}\\
\vspace{2.3cm}
Dimitri KUSNEZOV\footnote{Bitnet: DIMITRI\%NST@YALEVMS.}\\
\vspace{2mm}
{\em Center for Theoretical Physics, Sloane~Physics~Laboratory,}\\
{\em Yale University, New Haven, CT 06511-8167}\\

\vspace{1.3cm}

John SLOAN\footnote{Bitnet: SLOAN@OHSTPY.}\\
\vspace{2mm}
{\em Physics Department, The Ohio State University, Columbus, OH 43210}

\vspace{1.cm}

{\it July 1992}

\vspace{2.2cm}

\parbox{13.0cm}
{\begin{center}\large\sc ABSTRACT \end{center}
{\hspace*{0.3cm}
We investigate the use of global demons, a `canonical dynamics', as an approach
to simulating lattice regularized field theories. This deterministically
chaotic dynamics is non-local and non-Hamiltonian, and preserves the canonical
measure rather than $\delta(H-E)$. We apply this inexact dynamics to the 2D XY
model, comparing to various implementations of hybrid Monte Carlo,  focusing on
critical exponents and critical slowing down.  In addition, we discuss a scheme
for making energy non-conserving dynamical algorithms exact without the use of
a Metropolis hit.}}
\end{center}
\newpage
\begin{center}
{\large 1. Introduction}
\end{center}
\vspace{7mm}

One traditional dynamical approach to simulating ensemble averages has been
molecular dynamics (MD) algorithms.  In the simplest of these, micro-canonical
simulations, conjugate momenta are introduced for each degree of freedom in the
ensemble, and the resulting system is time-evolved according to Hamilton's
equations of motion.  The reversibility of Hamiltonian evolution then ensures
detailed balance, i.e. that the simulation is a Markov process. If  the system
is sufficiently large, and the interactions  sufficiently complex,  one usually
imposes the quasi-ergodic hypothesis, and hopes that the  simulation will
explore the desired ensemble.  Unfortunately, it is known that Hamiltonian
evolution conserves energy, and is therefore not ergodic.  In fact,
microcanonical simulations introduce an explicit factor of $\delta(H-E)$ into
the measure of the ensemble being simulated.  In order to use an MD algorithm
to obtain the correct ensemble, some additional method must be introduced to
integrate over the different energy surfaces $E$.

One method of dealing with this difficulty is embodied in  the hybrid molecular
dynamics (HMD) and hybrid Monte Carlo (HMC) algorithms\cite{hmca}--\cite{hmcc}.
In these algorithms, the micro-canonical equations of motion are integrated
along a `trajectory' for a time $T$, after which the momenta are touched with a
heat bath, changing the energy of the system.  As with any numerical
integration of Hamilton's equations, finite step size errors will build up
along the micro-canonical trajectories, leading to systematic errors in the
ensemble generated by HMD.  Although these errors can be controlled by making
the step size sufficiently small, this can become costly.  The HMC algorithm is
designed to correct for these $dt$ errors.  It does this by treating the
configuration at the end of a micro-canonical trajectory as a proposal for a
global update of the system, which is then accepted or rejected according to a
Metropolis hit.  If the equations of motion are reversible, this sequence of
configurations is then a Markov chain which, given ergodicity, is guaranteed to
produce the correct ensemble. The HMC and HMD algorithms are currently widely
used in lattice gauge theory simulations, especially in systems involving
dynamical fermions.  Typically, HMC is used in theories where the action can be
expressed as the volume integral of a local function, while HMD is used when it
cannot.

Although these algorithms are generally quite robust, they do have one
weakness, related to critical slowing down.  Associated with any observable $O$
is an autocorrelation `time' $\tau_O$, the time scale required for the
simulation to produce a statistically independent measurement of $O$. This
autocorrelation time will generally depend on the correlation length of the
system as a power law\cite{adk},
\begin{equation}
                                \tau_O =A\xi^z.\label{eq:texp}
\end{equation}
where $z$ is the dynamical critical exponent, and $A$ and $z$ will depend on
$O$. Critical slowing down occurs whenever $z>0$. $\tau_O$ represents the
typical amount of simulation time it takes for a local change in $O$ to
propagate across a correlated cluster. For multi-scale algorithms, such as
cluster algorithms, one can hope to obtain $z=0$, since these algorithms are
designed to change an entire  correlated cluster simultaneously. Unfortunately,
these algorithms are not easily generalized from one model to another, and have
not yet been implemented for most  lattice gauge theory models.  At the other
extreme, it is dangerously easy to obtain $z=2$ with an algorithm which
involves only local updates.  This just corresponds to diffusive transport of
fluctuations through the correlated cluster.  In principle, one should be able
to do no better than $z=1$ with a local algorithm, since the local algorithm
will restrict fluctuations to a finite propagation speed\cite{rutgers}.  One
can obtain  $z=1$\cite{adk,ktc} in HMC and HMD, but to do so requires a {\it
correlation length dependant tuning of the trajectory length.}  If the
trajectories are too short, then the frequent randomization of the momenta
causes the motion of the system through phase space to be diffusive (resulting
in a $z$ of $2$), while if the trajectories are too long the quality of
statistics goes down because energy conservation correlates the measurements
along each trajectory. In addition, the optimum trajectory length is likely to
differ for different  observables, forcing an inefficient trajectory length for
some of them. (This problem is especially severe for HMC, where only
one measurement is allowed per trajectory.  Running with trajectories twice as
long as is necessary is therefore equivalent to a factor of two slow-down in
the code.)

We would like to contrast these hybrid algorithms based on micro-canonical
evolution to a purely dynamical approach we term {\sl global demons}. Like the
hybrid algorithms, global demons are an easily implemented, very general
approach to simulating ensemble averages. Unlike the hybrid algorithms, the
global demon formulation has nothing to do with Hamiltonian dynamics. For
example, it can be defined completely in terms of coordinates alone if so
desired: the presence or absence of a symplectic structure is irrelevant. This
should be contrasted to the MD based algorithms, where a partition function
whose action depends only on coordinates is usually augmented to include
fictitious conjugate momenta in order to define a Hamiltonian or Poisson
structure.  Another way to say this is that while evolution using Hamiltonian
dynamics generates a microcanonical ensemble, evolution in the global demon
system generates a canonical ensemble. The global demon equations of motion are
deterministic and time reversal invariant, and are  designed to evolve through
the physically accessible regions of configuration space (it is {\it not}  a
phase space) such that the trajectory fills configuration space with a density
reproducing the correct ensemble.  Consider, as an illustration, 1000
points in a phase space $(q,p)$ which lie equally spaced on a unit circle, as
shown in the left column of Fig.~1, at  $t=0$. The points are connected in
order to see how neighboring points behave. Under microcanonical evolution of
the 1-d harmonic oscillator equations of motion ($t=1$ is the natural time
scale of the dynamics),
\begin{equation}
         \dot{q} = p,\qquad \dot{p} = -q \; ,
\end{equation}
this circle will be preserved, and the points will rotate, preserving the
figure at all later times. In contrast to Hamilton's dynamics,  global demon
dynamics will result in a rapid dissemination of neighboring points through the
space. The time evolution of the circle for a 1-d harmonic oscillator
Hamiltonian $H=(p^2+q^2)/2$ is shown at time $t=10$ and $t=20$, in our
canonical dynamics.  What is striking is the speed at which neighboring points
on the circle evolve to opposite sides of the the space. Even at $t=20$, one
can see that the phase space density is nearing the desired ensemble
$\exp(-\beta H)$. In the right column of Fig.~1, we have the same
situation for an $SU(2)$ Hamiltonian $H=J_z^2/2$, whose phase space,
parameterized by $(J_x,J_y,J_z)$, is the unit sphere. Here an initial condition
of a circle at $J_z=0.5$ is evolved in a similar manner.  Again, the rapid
divergence of neighboring points is striking. Since global demons generate a
deterministic chaotic dynamics, the danger of diffusive motion through phase
space present in the hybrid and other stochastic algorithms is absent here. The
one drawback to the global demons approach is that we have been unable to
determine a way of making it exact, i.e. removing the $dt$ errors in the
ensemble arising from finite step size.

In this article, we investigate an application of the global demon algorithm to
a lattice field theory. We are interested in understanding the critical
properties of the dynamics near phase transitions, and how tuning the dynamics
can improve convergence. We also want to examine the correctness of
measurements in this inexact dynamics, as well as ways to make it exact without
recourse to stochastic techniques. We choose the 2D XY model, since it has been
well studied in the past in a variety of algorithms, and has an infinite order
phase transition around which we can investigate the manifestations of critical
slowing down.  In section 2, we present the global demon dynamics for
unconstrained systems. The implementation of this dynamics  for the XY model in
section 3 explores the behavior of simulations as various  parameters of the
algorithm are tuned. We find that the algorithm is quite robust, obtaining good
results with little tuning.  In addition, we compare the critical behavior of
the algorithm to several implementations of HMC. We conclude in section 4.
Finally, in an appendix, we present a scheme which, in principle, should remove
the finite $dt$ errors from the algorithm dynamically. Although our present
implementation of the global demon algorithm is not exact, we have chosen to
compare our results to HMC, rather than HMD.  The main reason for this choice
is that we are treating the HMC results as a control, and would like them to be
as free of systematic errors as possible.  Since the HMC and HMD algorithms are
so similar, however, it is quite likely that qualitative conclusions about the
critical behavior of HMC will also be correct for HMD.

\vspace{1cm}
\begin{center}
{\large 2. Global Demon Dynamics}
\end{center}
\vspace{1cm}

Let us consider a system characterized by an action $S(x)$ and
coordinates $x=(x_1,...,x_n)$. The ensemble averages of this system will have
the generic form
\begin{equation}
 \VEV{\co}= \frac{1}{ Z}\int\;{\cal D}\mu(x)\;e^{-\beta S(x)}\co,\label{eq:ens}
\end{equation}
where
\begin{equation}
   Z = \int\; {\cal D}\mu(x)\; e^{-\beta S(x)},
\end{equation}
is the normalization, and the measure ${\cal D}\mu(x)$ might include
constraints (for example, symmetries associated with a Lie algebra). In this
article, we are only concerned with the situation when the measure is trivial,
${\cal D}x$. If the variables $x=(q,p)$ include canonically conjugate
coordinates and momenta,
$S(x)$ can be taken as a Hamiltonian: $H(p,q)=S(x)$. Otherwise,  in what is
more or less standard practice, conjugate momenta are introduced and added to
$S(x)$ to make the exponent in Eq.~(\ref{eq:ens}) resemble a  Hamiltonian
\begin{equation}
   H(p,x) = \frac{1}{ 2}\sum_{i=1}^n p_i^2 + S(x),
\end{equation}
and the measure is modified to ${\cal D}x{\cal D}p\exp(-\beta H)$, with
appropriate normalization. Molecular dynamics is now easily
implemented,  leading to the equation of motion
\begin{equation}
   \dot{x_i} = p_i\, ,\qquad \dot{p_i} = -\partder{S(x)}{x_i},\qquad
                               (i=1,...,n).\label{eq:mcd}
\end{equation}
This dynamics, combined with momentum refreshes and
global metropolis hits, produce the HMD and HMC algorithms.

Let us now pass to the canonical dynamics for such a
system\cite{cmda}--\cite{cmdc}.
In contrast to the microcanonical dynamics, the energy and the symplectic
structure are no longer preserved. Rather, the measure itself is preserved
directly by the dynamics. We can define such dynamics in many
ways. For instance, instead of Eq.~(\ref{eq:mcd}), we could take
\setcounter{equation}{0}
\setcounter{section}{7}
\renewcommand{\theequation}{\arabic{section}\alph{equation}}
\begin{equation}
    \dot{x}_i = - \kappa\, \frac{\beta}{ n}\, \frac{dG(w)}{ dw}\;
      F_{i}(x)\, ,\qquad (i=1,...,n).
\end{equation}
Here $G(w)$ and $\fax$ are arbitrary functions of a global demon variable
$w$ and coordinates $x$, respectively, and  $\ka$ is a coupling constant. The
number of global demon interactions (the right hand side of (7a)) is
unrestricted.  In this example we have used 1, while 2 or 3 are usually
sufficient, regardless of $n$.  This type of treatment can be
viewed as a deterministic version of Parisi and Wu's stochastic
quantization\cite{pwu,cmdb}. An alternative formulation of the $x$ dynamics
which includes a relic of the underlying Hamiltonian $H(p,x)$ is
\begin{eqnarray}
 \dot{x}_i&= &p_i - \frac{\kappa_1\beta}{ n}\; \frac{dG_1}{dw_1}\; F_{1i}(x),\\
 \dot{p}_i &= &-\partder{S(x)}{x_i} - \frac{\kappa_2\beta}{ n}\;
                      \frac{dG_2}{dw_2}\; F_{2i}(p).
\end{eqnarray}
\setcounter{equation}{7}
\renewcommand{\theequation}{\arabic{equation}}
An important observation here is that while we can retain a Hamiltonian
sub-structure to the dynamics, it is not responsible for the ergodicity in the
full configuration space, and can be retained or altogether removed. This will
have some effect on the convergence, since the Hamiltonian forces can provide
additional decorrelation. In non-equilibrium simulations, it is more convenient
to use (7b-c)  since there is a closer link to the thermodynamics of
$H(p,x)$\cite{cmdd}. Eqs. (7b-c) also have a microcanonical limit when
$\kappa_\alpha=0$. Since we are going to compare the global demon approach to
HMC, we retain the momenta to have a greater parity between the two algorithms.

With the introduction of the global demons $w_\alpha$, a larger configuration
space $\{\phi\}$ must be defined, where $\phi=(x_1,...,p_n,w_1,...w_m)$ . In
$\phi-$space we can  define a new action $f$, which is determined by the
equations of motion (7)  in the following way:
\begin{equation}
  f(x,p,w) = S(x) + \frac{1}{ 2}\sum_{i=1}^n p_i^2 + \sum_{\alpha=1}^m
  G_\alpha(w_\alpha),\qquad\quad \rho_f=e^{-\beta f}.\label{eq:meas}
\end{equation}
Unlike $H(p,x)$ in molecular dynamics, $f$ is {\sl not} preserved by the
equations of motion. While the definition of $f$ is not unique (in the sense
that the measure for the variables $w$ is arbitrary), it is natural as well as
convenient in determining the dynamics of the global demons, by providing a
particular solution to the continuity equation below. Eq.~(\ref{eq:meas})
defines the $\phi-$space measure as ${\cal D}x{\cal D}p{\cal D}w\exp(-\beta
f)=\rho_f\; {\cal D}\phi$. The global demon dynamics can then be determined by
requiring that  $\rho_f$ be a stationary solution of a generalized Liouville
(continuity) equation in configuration space:
\begin{equation}
  0 = \partder{\rho_f}{t} + \sum_{i=1}^{n+m}
                   \partder{(\dot{\phi}_i\rho_f)}{\phi_i}.\label{eq:prob}
\end{equation}
This is equivalent to requiring that the master equation, enforcing
conservation of probability under evolution of the ensemble, be satisfied.

The equations of motion for the demons are now found by requiring that they,
combined with the generalized dynamics of (7), satisfy Eq.~(\ref{eq:prob}). A
direct substitution of $\rho_f$ and $\dot{\phi}_i$ (Eqs. (7b-c) ) into
Eq.~(\ref{eq:prob}) allows one to solve for $\dot{w}$:
\begin{eqnarray}
   \dot{w}_1 &= &\frac{\kappa_1}{ n} \left( \beta F_{1i}
               \partder{S}{x_i} - \partder{F_{1i}}{x_i}\right),\label{eq:grd}\\
   \dot{w}_2 &=& \frac{\kappa_2}{ n}
               \left( \beta F_{2i} p_i - \partder{F_{2i}}{p_i}\right).\nonumber
\end{eqnarray}
If we had chosen to neglect the momenta and used (7a), the form of
Eq.~(\ref{eq:grd}) would be unchanged. Eqs. (7) and ~(\ref{eq:grd}) define a
dynamics which by construction preserves the measure Eq.~(\ref{eq:meas}). (It
is worth noting that while we have taken an exponential form for the density,
in general we can take an arbitrary function $\rho$ and still use this same
procedure.)

Microcanonical dynamics preserve the phase space volume exactly, since the
divergence of the equations of motion,
\begin{equation}
    \partder{\dot{q}_i}{q_i} + \partder{\dot{p}_i}{p_i},
\end{equation}
trivially vanishes by Hamilton's equations of motion. The  global demon
equations of motion (7),~(\ref{eq:grd}), on the other hand, do allow for
fluctuations in the $\phi$-space volume, which can be quite
large. Writing these equations as $\dot{\phi}_i = {\cal F}_i(\phi)$, the
divergence is explicitly
\begin{equation}
  \partder{\dot{\phi}_i}{\phi_i} =
     -\frac{\beta}{ n}\left(\kappa_1\frac{dG_1}{ dw_1}
          \partder{F_{1i}}{x_i} + \kappa_2
           \frac{dG_2}{dw_2}\partder{F_{2i}}{p_i}\right).\label{eq:volf}
\end{equation}
This local `breathing' of $\phi-$space is controlled by the arbitrary
functions $G$
and $F$. Although this behavior is not microcanonical, there is nevertheless
an invariant quantity, called the pseudoenergy ${\cal E}$, which is preserved:
\begin{equation}
   {\cal E} = f(x,p,w) + \frac{1}{\beta}\int_0^t dt'\;
   \partder{\dot{\phi}_i}{\phi_i} .\label{eq:pse}
\end{equation}
One can check directly that $\dot{\cal E}=0$.

There is clearly some freedom in defining the dynamics: the functions $G$ and
$F_i$ and the coupling strength $\ka$. The only restriction on $G(w)$ is that
the measure Eq.~(\ref{eq:meas}) is normalizable; in general the auxiliary
variables $w$ can have any desired measure. In practice, highly non-linear
functions are impractical since they will require small integration time steps.
For these reasons, it is convenient to take $G=w^2/2$ or $G=w^4/4$. A necessary
condition for $F_{i}$ is for it to be at least linear in its argument, the
minimal requirement for the existence of the fluctuations in the volume
{}~(\ref{eq:volf}). The precise relation to the fluctuations in a volume $V$,
or
equivalently, the instantaneous $\phi-$space compressibility, can be found
using the divergence theorem
\begin{equation}
     \frac{dV}{ dt} =-\frac{\beta}{ n}\int_V {\cal D}\phi\;
           \left(\kappa_1\frac{dG_1}{ dw_1} \partder{F_{1i}}{x_i} + \kappa_2
           \frac{dG_2}{dw_2}\partder{F_{2i}}{p_i}\right).
\end{equation}
In this paper, we do not explore the effect of different choices of $G_i$,
$F_i$.  Such studies have been done on smaller
systems\cite{cmda}--\cite{cmdc}.

Finally, we observe that the equations of motion (7), (\ref{eq:grd}) will have
no stable fixed points\cite{orbit}. This is the case since the sum of the
Lyapunov exponents is related to the average rate of change of total volume of
$\phi-$space. By the Liouville equation, this will necessarily
vanish\cite{billa}:
\begin{equation}
   \sum_{i=1}^{n+m}\lambda_i = \VEV{\partder{\dot{\phi}_i}{\phi_i}}=0.
\end{equation}
%

\vspace{1cm}
\begin{center}
{\large 3. Implementation for the 2D XY Model}
\end{center}
\vspace{1cm}

The 2D XY model consists of spins located on the sites of a two
dimensional square lattice, which are free to rotate in the plane. The action
is given by
\begin{equation}
        V(\theta) = -\sum_{<ij>}  Re U_i U_j^\dagger
                  = -\sum_{<ij>} \cos{(\theta_i-\theta_j)},
\end{equation}
where the sum is over nearest neighbors, and the $U_j\equiv e^{i\theta_j}$ are
elements of $U(1)$ located at each lattice site $j$. In two dimensions, this
model exhibits a Kosterliz-Thouless phase transition near $\beta\sim
1$\cite{kta}--\cite{kte}. Above the phase transition, the dynamics is dominated
by dissociated vortex-antivortex pairs. These pairs become tightly bound below
the phase transition, where the dynamics is dominated by spin waves. The K-T
phase transition is infinite order, characterized by an exponentially diverging
correlation length ($\xi$):
\begin{equation}
    \xi = a_\xi\exp( b_\xi (T-T_c)^{-\nu}).
\end{equation}
Numerical simulations indicate similar critical behavior for finite
lattices\cite{ktd,kte}. Near the critical temperature, the system
experiences an
exponential increase in the correlation length $\xi$, which can lead to
critical slowing down in simulations by virtue of Eq.~(\ref{eq:texp}).

\vspace{7mm}
\noindent{\sc 3.1 Equations of Motion}
\vspace{7mm}

The implementation of global demons to the XY model is straight forward.
Following the equations of motion (7b-c) and ~(\ref{eq:grd}), we have
\begin{eqnarray}
     \dot{\theta}_i & = & p_i - \frac{\kappa_2\beta}{ n}w_2\sin^3\theta_i
                 ,\qquad (i=1,...,n),\label{eq:eoma}\\
     \dot{p}_i      & = &-\partder{V(\theta)}{\theta_i} -
                        \frac{\kappa_1\beta}{ n}w_1^3 p_i.\nonumber
\end{eqnarray}
$F_i(\theta) = \sin^3\theta_i$ is chosen to respect the  periodicity in
$\theta$, while $G_i(p)=p_i$ has no such restriction. This choice was motivated
only by simplicity, and in general, we could take more complicated
interactions, and include additional global demons. The corresponding equations
for the global demons are then
\begin{eqnarray}
        \dot{w}_1 & = &\kappa_1\left[ \frac{\beta}{ n}\sum_i p_i^2 -
                               1\right],\label{eq:eomb}\\
        \dot{w}_2 & = &\frac{\kappa_2}{ n}\left[
              \beta\sum_i\partder{V(\theta_i)}{\theta_i}\sin^3\theta_i
                - 3\sum_i \sin^3\theta_i\cos\theta_i\right].\nonumber
\end{eqnarray}
We used leapfrog integration, which included a Taylor expansion so that the
$O(dt^2)$ errors in a time step cancel. The pairs $q$, $w_1$ and $p$, $w_2$
were updated in alternate steps. $w_1$ was taken with $q$ since they both
involve momenta, and $w_2$ with $p$ since they both involve coordinates.
Our general studies of the
systematics of the model under tuning of parameters were performed on a  $16^2$
lattice, while the studies of the critical exponents were done on a $64^2$
lattice, to allow longer correlation lengths. The particular choice of
functions $F$ leads to a small non-ergodicity for this particular system: the
momentum zero mode cannot change sign. (We have corrected for this by
occasionally (every 64 trajectories) refreshing the  momenta, using the same
procedure as in HMD.  In general, this is probably a good idea, to ensure that
the evolution is ergodic.  This particular non-ergodicity
could also have been corrected by  a small modification of the equations of
motion.) We have verified that the equations of motion are correct to $O(dt^3)$
on a  time step, leading to  $O(dt^2)$ systematic errors in observables, by
computing the  $dt$ behavior of  several observables in Fig.~2,
demonstrating the quadratic behavior of the systematic error.

\vspace{7mm}
\noindent{\sc 3.2 Hybrid Monte Carlo}
\vspace{7mm}

We have used the critical properties of HMC as a benchmark for comparison of
our global demon approach, studying three of its variations\cite{ktc}. The
equations of motion are the same as those used for global demons, except with
$\kappa_i=0$. By modifying the length of the HMC trajectory between momentum
refreshes, we modify the decorrelation time\cite{hmcb}. The first variation,
denoted  HMC-1, has trajectories of length 1, where the highest frequency of
the free  theory is $(2\pi)^{-1}$. While this `standard' choice is easy to
implement,  it suffers from severe critical slowing down, with $z=2$ in
Eq.~(\ref{eq:texp}).  The critical behavior should be improved by choosing the
trajectory length proportional to the  spatial correlation length $\xi$ of the
system\cite{adk}.   The two variations we consider are denoted HMC-S, for
$T=\xi$, and HMC-L, for  $T=2\pi\xi$. Again we point out that, in order to make
this choice, we require the very information which we are attempting to
measure. We were fortunate to have previous results for $\xi$ available to
us\cite{kte}, but in general this is not likely to be the case.

The integration time step was kept fixed at $dt=.1$ along trajectories of
length $T$. This value of $dt$ was chosen so that the acceptance rate of the
global Metropolis hit in the HMC algorithm was approximately $80\%$.   For HMC,
it is necessary to use random trajectory lengths for optimum
relaxation\cite{hmcb}: we chose $T$ uniformly distributed  on the interval
$(.5\VEV{T}, 1.5\VEV{T}]$, with $\VEV{T} = 1$ for HMC-1, and
$\VEV{T}\approx\xi, 2\pi\xi$ for HMC-S,L.  We also made several runs of HMC-L
using exponentially distributed random trajectory
lengths.  We found that this does not lead to any improvement over the runs
with uniformly distributed trajectories, and may even have resulted in
slightly noisier measurements.  One interesting difference between HMC and
global demons is that, for global demons, $T$ simply denotes time between
measurements along a single trajectory - the evolution of the simulation is
completely unaffected by the choice of $T$.
At the end of a HMC trajectory we performed a global
metropolis hit, after which we performed measurements and refreshed the momenta
by choosing new, gaussian distributed $p_i$.

\vspace{7mm}
\noindent{\sc 3.3 Coupling Strength Dependence}
\vspace{7mm}

In the micro-canonical algorithms HMC and HMD, the rate at which the simulation
covers phase space in the non-ergodic directions (i.e. changes energy) is
controlled by the time between momentum refreshes.  If the trajectories are too
long, the system changes total energy very slowly, leading to autocorrelations
on timescales proportional to the trajectory length.  If the trajectories are
too short, on the other hand, the motion between energy shells is rapid, but
motion in the micro-canonical directions is diffusive, leading to a dynamical
critical exponent of $2$.  This means that, for  large correlation lengths, the
efficiency of the algorithm can vary as a power of $T/T_{opt}$, where $T_{opt}$
is the optimum trajectory length. As shown in Refs.~\cite{hmcb,adk}, $T_{opt}$
should be proportional to the correlation length of the system, which is not
known {\it a priori}.  Thus it is necessary  to perform a sensitive tuning
which depends on a parameter measured in  the simulation.

In the global demon algorithm, on the other hand, the parameter which controls
energy (or action) non-conservation is just the coupling $\kappa$ of the demons
to the system.  In the limit $\kappa \rightarrow 0$, the demons decouple and
ergodicity is lost.  If $\kappa$ becomes too large,  the equations of motion
will suffer the characteristic instabilities of discretized dynamics. In
contrast to the HMC and HMD algorithms, however, we can make an {\it a priori}
choice of $\kappa$ which works quite well at all values of the correlation
length. To do this, consider the change in total action, $\Delta S =
S(\phi_2)-S(\phi_1)$, in a single time step,
\begin{equation}
    \Delta S \simeq \Delta t\;
      \partder{S}{\phi_i}\dot{\phi}_i\quad ,\label{eq:nab}
\end{equation}
where $S$ now refers to the total global demon action $f$ in
Eq.~(\ref{eq:meas}).  Using equations (7)-(\ref{eq:grd}) and ~(\ref{eq:nab}),
we see that $<|dS/dt|>$ is proportional to $\kappa$, with a constant of
proportionality of order one. Thus, the change in $S$ along a trajetory of
length $T$ should be $\Delta S \simeq \kappa T$.  To set the scale of $\Delta
S$,  we can compute its expectation value if two consecutive measurements are
totally decorrelated:
\begin{equation}
      \sigma = \sqrt{\langle\Delta S^2\rangle}
             = \sqrt{2(\langle S(\phi)^2\rangle -\langle
          S(\phi)\rangle^2 )} = \frac{1}{\beta}\sqrt{2nC_s}\quad ,
\end{equation}
where $C_s$ is the specific heat of the system. In order for the action to
decorrelate between measurements (with $T=1$), we  conclude that the optimal
choice of $\kappa$ is likely to be $\kappa \approx {\cal O}(\sqrt{n})$.  Note
that this philosophy of forcing large fluctuations in energy along a trajectory
is inherently different from HMC, where (in order to avoid prohibitively low
acceptance rates) $\Delta S$ is ${\cal O}(1)$.

To investigate the behavior of the global demons algorithm under tuning, we ran
a series of simulations on a $16^2$ lattice. (Except where otherwise indicated,
we used an integration time step of $dt=0.1$ and measuring at intervals
$T=1=10dt$; we will call this time between measurements a trajectory length,
even though no momentum refresh is performed).  Along the trajectory, we
determine the square of the change in action between measurements, $\Delta S^2
= [S(t)-S(t-T)]^2$,  which is then averaged along the entire trajectory to
obtain $\sqrt{\VEV{\Delta S^2}}$.  The result gives a guide as to how fast the
trajectory can diffuse through configuration space. In Fig.~3, we plot the
quantity $\overline{\Delta S}$, which we call the diffusiveness, defined by
\begin{equation}
     \overline{\Delta S} = \frac{ \sqrt{\VEV{\Delta S^2}}}{ \sigma},
\end{equation}
as a function of coupling strength $\kappa$, for simulations at  representative
values of $\beta$ both above and below the phase transition. In the limit
$\kappa_i=0$, the equations of motion ~(\ref{eq:eoma})-(\ref{eq:eomb}) are
microcanonical and $S$ is preserved, as indicated in the figure. For small
couplings, the microcanonical component of the dynamics is only slightly
perturbed by the canonical component, and the ergodicity is weak. The
convergence times here are quite large\cite{cho}. For
$\kappa_i\sim\sqrt{2n}\sim 23$, the value of
$\overline{\Delta S}$ can be seen to saturate near unity, the value expected
when two consecutive measurements are uncorrelated; here the steps are quite
large through configuration space. For larger values of the coupling, the
change in action remains saturated. Convergence is also generally slower for
larger $\kappa$, since the additional decorrelation produced by the
microcanonical component to the  dynamics is reduced. The reduction of
$\overline{\Delta S}$ in Fig.~3 for $\beta\sim 1$ can be attributed to
critical slowing down. However, while the correlation lengths become quite
large, the dip in $\overline{\Delta  S}$ is not so noticeable. {\sl In this
respect, critical slowing down does not seem to hinder the dynamics, nor does
it require any special tuning of $\kappa$.}

In Fig.~4, the $\beta$ dependence of $\overline{\Delta S}$ is plotted for
simulations at fixed coupling strengths. By selecting the $\kappa = 32$ curve,
for instance, we see that we can study both the low and high temperature
properties of the $XY$ model, as well as the phase transition, without
modifying $\kappa$. While there is a small dip in the curves near $\beta\sim
1$, critical slowing down does not seem to strongly effect this measure of the
dynamics as one approaches the phase transition from either side. An important
result is that the couplings $\kappa$ are essentially independent of $\beta$ as
well as the details of the physics of the model under study.

It should be emphasized that the runs with $\kappa\ll\sqrt{2n}$ converge slower
as $\ka$ decreases, and ultimately do not converge for the microcanonical limit
$\kappa=0$. We have checked the convergence of the dynamics to the proper
ensemble by measuring and subsequently histogramming $\wa$ and $p_i$ along the
trajectory and comparing them to their exact analytic distributions, finding
that convergence is best above $\kappa\sim \sqrt{2n}$. We can also examine the
$\kappa$ dependence of measurements at fixed $\beta$, illustrated in Fig.~5 for
$\beta=1/1.1$. What we see is that the measurements for coupling strengths
roughly 5-8 times saturation do not exhibit any systematic deviation as
$\kappa$ increases. For much larger $\kappa$ (at fixed $dt$), there will be the
characteristic instabilities associated with difference equations. However, the
value $\kappa=\sqrt{2n}$ clearly is not  near this instability limit, and we
can safely use it. In the low $\kappa$ limit, we are close to micro-canonical
dynamics, and $S$ begins to become  approximately conserved. This slow
diffusion results in long time correlations and poor statistics. This figure is
typical of other temperatures, above and below the phase transition.

\vspace{7mm}
\noindent{\sc 3.4 Choice of Trajectory Length (i.e. Measurement Frequency)}
\vspace{7mm}

An indication of how rapidly measurements decorrelate is shown in Fig. 6.
There we plot the diffusiveness $\overline {\Delta S}$ as a function of
trajectory length $T$, for $\kappa_1=\kappa_2=16$ and $\beta=1$. We observe a
saturation in the trajectory length near $T=2$.  Because measurements do not
effect the time evolution of  the global demon trajectory (they are not
associated with any  Metropolis hit or momentum refresh), the choice of
frequency of measurements is governed by the relative costs of the time
evolution and measurement routines.  Because our measurement algorithm was
relatively inexpensive, we chose to use $T=1$.

\vspace{7mm}
\noindent{\sc 3.5 Observables}
\vspace{7mm}

The observables we measured include the energy $E$, lattice magnetization $M$,
topological charge $Q$, defined by
\begin{eqnarray}
       E  & =& -\frac{1}{ n}\sum_{<ij>}Re U_iU_j^\dagger,\nonumber\\
       M  & =& \frac{1}{ n} \sum_i U_i,\\
       Q  & =& \frac{1}{ n} \sum_p q_p.\nonumber
\end{eqnarray}
The sum in $Q$ indicates the sum over all plaquettes of the number of
positive topological charges occupying that plaquette.  (For an exact
definition of the topological charge, see e.g.\cite{hands}.)
Corresponding to these observables, we can define the specific heat and
susceptibilities:
\begin{eqnarray}
          C_v   &=& \beta^2 n (\VEV{E^2}-\VEV{E}^2),\nonumber\\
          \chi_Q&=& n (\VEV{Q^2}-\VEV{Q}^2),\\
          \chi_M&=& n (\VEV{ (ReM)^2} + \VEV{(ImM)^2}).\nonumber
\end{eqnarray}
In both our global demon and HMC runs, we started with about $20000/T$
trajectories for thermalization, followed by $160,000/T$ trajectories of data,
where $T$ is the trajectory length. Statistical errors in observables were
obtained by binning measurements in bins of size $2^n$. The errors quoted in
our tables use the smallest bin larger than $8\tau_M$, where $\tau_M$ is the
integrated autocorrelation time of the total magnetization.  The errors in the
susceptibilities were obtained from the errors in the corresponding observables
by assuming gaussian fluctuations on a timescale $\tau_{\cal O}$, where
${\cal O}$ is the appropriate observable.
A selection of our observables are indicated in Table 1. We find that the
global demon results usually agree with the HMC results within a few $\sigma$,
which indicates that the systematic errors are not large. They could, of
course, be further reduced by extrapolating to $dt=0.$

\vspace{7mm}
\noindent{\sc 3.6 Autocorrelation Functions and Decorrelation Times}
\vspace{7mm}

Because we have a dynamical algorithm, the trajectory has a memory, which will
be reflected in the auto-correlation functions. This can be analyzed by
examining auto-correlation functions of the observables $M$, $E$, $Q$, and $S$.
When the couplings $\kappa$ are small, the dynamics is near the microcanonical
limit, and decorrelation is very poor. Typical auto-correlation functions for
$\ka_1=\ka_2=1$ are shown in Fig.~7 (dots) for $\beta=1/1.1$. Here we define
\begin{equation}
  \delta {\cal O}(t) = {\cal O}(t)-\VEV{\cal O},
\end{equation}
as the fluctuation from the mean. As the couplings are increased near their
optimum values, the ringing disappears, and decorrelation times become better
defined quantities, indicated by the solid curve with $\kappa_1=\kappa_2=16$.
The corresponding HMC autocorrelation functions are shown as well.
Parenthetically, this type of ringing can also occur in HMC simulations if one
uses a constant trajectory length. The comparison to our HMC runs at this
$\beta$ is shown in Fig.~8.

A comparison of auto-correlation functions for the total lattice magnetization
$M$ for global demons to the various implementations of HMC are shown in Fig.~9
for a selection of temperatures. It is clear that HMC does not significantly
out perform global demons in terms of decorrelation, no matter how `optimal'
the trajectory length. The points in these curves indicate the actual number of
data points. Hence while the number of global demon measurements is given by
$t$, the optimal HMC runs have between one and two orders of  magnitude smaller
sampling rate in order to have similar decorrelation behavior.

\vspace{7mm}
\noindent{\sc 3.7 Critical Exponents}
\vspace{7mm}

The integrated autocorrelation times $\tau$ are defined for a given quantity
${\cal O}$ as
\begin{equation}
  \tau_{\cal O} = T\left[\frac{1}{2} + \sum_{t=1}^\infty
  \frac{\VEV{\delta{\cal O}(0)\delta{\cal O}(t)} }
       {\VEV{\delta{\cal O}(0)\delta{\cal O}(0)} }\right].
\end{equation}
Note that we are measuring $\tau$ in units of total time evolved
rather than number of trajectories. In Fig.~10 we present a ln-ln
plot of the decorrelation times $\tau$ vs. the correlation length $\xi$ for
several observables. The fit parameters are indicated in Table 2, and are only
indicative of the critical behavior, since they will depend strongly on
systematic effects. What is seen is that in almost every case, the global
demon prefactor and critical exponent are smaller than the HMC results. The
integrated autocorrelation times are tabulated in Table 3. In Ref. \cite{adk},
it was argued that the critical exponent is unity when $T$ is proportional to
the lowest frequency mode in the system for a free field theory. The results in
Tables 2--3 seem to be good evidence that their results  are qualitatively
correct in an interacting field theory as well.

The critical behavior of global demon dynamics will also depend on the coupling
strengths. In Fig.~11 we plot $\tau$ as a function of $\kappa$ for
simulations at several values of $\beta$, on a $16^2$ lattice. As $\kappa$
increases, the system tend to decorrelate faster, again generally saturating
above $\kappa\sim\sqrt{2n}$.  In Fig.~12, the $\beta$ dependence of
simulations at $\kappa=1,4,16,64$ are indicated. The $\kappa=1$ runs have the
highest decorrelation times as expected, but we also observe that the high
temperature phase is rather insensitive to the value of the coupling. Although
convergence of the trajectory to the correct ensemble will always depend
strongly on the coupling strength, the decorrelation times of both weakly
ergodic and strongly ergodic trajectories are very similar. The effects of
critical slowing down are particularly noticeable in $\tau_Q$ and $\tau_E$ near
$\beta\lsim 1$. The peak in the $\tau_E$ is closely related to the dip in
$\overline{\Delta S}$ in Fig.~4.
The reason is that the diffusiveness
measures the maximum rate at which the total energy $S$ can change, so if
$\overline{\Delta S}\ll 1$, the potential energy $E$ will change slowly.

One might conclude from Figs.~11--12 that the coupling strength
dependence is not too important, and that $\kappa\sim 2$ is roughly equivalent
to $128$. Clearly, while the decorrelation times are indicative of the
dynamics, they do not provide the complete picture of the situation. For
example, information such as the ringing in the autocorrelation functions
(see Fig.~7) average out, and are not strongly reflected in the value
of $\tau$. We also see that simulations with very similar decorrelation times
can have disparate values of the diffusiveness $\overline{\Delta S}$.
But in all these guides, the tuning is consistently optimal for
$\kappa\sim\sqrt{2n}$.

\vspace{1cm}
\begin{center}
{\large 4. Conclusions}
\end{center}
\vspace{1cm}

We have studied the global demon dynamical approach to simulating lattice
regularized field theories. This method breaks away from the conventional
Hamiltonian wisdom, defining a deterministically chaotic, time reversal
invariant `canonical' dynamics which rapidly fills configuration space with the
desired ensemble. We have taken a particularly simple implementation of global
demons using two coupling functions and examined its critical behavior,
comparing to HMC using various trajectory lengths.

We have found that the algorithm is very stable under tuning of the various
parameters. In particular, once $\kappa$ is large enough, the quality of the
results seem to be independent of $\kappa$ until the simulation becomes
unstable. It appears that $\kappa\approx\sqrt{2n}$ is a good rule of thumb for
which the simulation will perform well in all regimes. In addition we found
that the systematic errors were small (a few standard deviations), and that the
critical slowing down properties of the algorithm were competitive with or
better than the best implementations of HMC. The main fault with the algorithm
is that it is not exact, i.e. there is a systematic error associated with the
numerical integration. This problem is addressed in the appendix.

One advantage of this approach is that there is no barrier in principle to
obtaining $z<1$ (note our estimated critical exponents for $S$, $E$ and $Q$ in
Table 2). In contrast, local algorithms such as HMC and HMD are limited by
$z\sim 1$\cite{rutgers}. In addition, the dynamical nature of the algorithm
has allowed extensions to non-equilibrium situations\cite{cmdd}.  One
possible improvement which we have not examined is Fourier acceleration. This
technique improves $z$ to $0$ for HMC in free field theory, and may help our
approach.

The numerical computations in this work were performed on the Cray {Y-MP8/864}
at the Ohio Supercomputer Center.  We thank Aurel Bulgac, Robert Edwards, Rajan
Gupta, Bill Hoover, Tony Kennedy, Greg Kilcup, Klaus Pinn, Junko Shigemitsu and
Beth Thacker for useful conversations.  This work was supported under  DOE
grants DE-AC02-ER01545 and DE-FG02-91ER40608.

\newpage

\begin{center}
{\large Appendix: Exact Dynamics?}
\end{center}
\vspace{1cm}

\setcounter{equation}{0}
\setcounter{section}{1}
\renewcommand{\theequation}{\Alph{section}.\arabic{equation}}

One major weakness of the global demon algorithm presented in this paper is the
systematic error associated with finite step size.  In HMC, this error is
eliminated by performing a global metropolis hit before every  measurement.
Unfortunately, this technique has no hope of working for  an algorithm with
global demons, since the trajectories do not conserve energy. One might hope to
use the pseudo-energy ${\cal E}$ in this capacity, since it is conserved.
However, the memory term in Eq.~(\ref{eq:pse}) precludes this. For Hamiltonian
systems, one can implement symplectic integrators\cite{yosh} to render the
dynamics exact, or one can introduce a global Metropolis hit as in HMC.  For
our non-Hamiltonian `canonical' dynamics, this procedure is not so clear.
Although we have been unable to satisfactorily solve this problem, we mention
here one approach which we have tried.  We present this method because,
although it is presently numerically impractical, it appears to be correct in
principle. In addition, it should be applicable to any dynamical simulation
which does not conserve energy, and is quite different from the traditional
method of using metropolis hits to ensure exactness.

Consider the exact (i.e. $dt=0)$ equations of motion, $\dot{\phi} = F(\phi)$,
which, by construction, preserve the measure $\rho(\phi)=\exp(-\beta f(\phi))$.
When we discretize time, $\rho$ is no longer preserved exactly. Let us assume
that there exists some measure, $\tilde \rho\not= \rho$, which is preserved by
the discretized equations of motion $\phi_{n+1} = M(\phi_n)$, where $M$ is the
time evolution operator. We can now define a correction factor $\alpha$ such
that, up to normalization,
\begin{equation}
   \rho(\phi) = \alpha(\phi) \tilde \rho(\phi).
\end{equation}
If we know $\alpha$, we can obtain the exact expectation values of observables
through convolution:
\begin{equation}
       \VEV{\cal O} = \frac{ \int {\cal O}\rho d\phi }{ \int \rho d\phi }
		    = \frac{\int ({\cal O}\alpha ) {\tilde \rho} d\phi}{
		       \int \alpha {\tilde \rho} d\phi} = \frac{\VEV{ {\cal
		       O}\alpha}'}{\VEV{\alpha}'},
\end{equation}
where the prime indicates evaluation in the $\tilde \rho$ ensemble.
The discretized Liouville (continuity) equation tells us that after one
integration time step, we preserve the measure $\tilde \rho$:
\begin{equation}
  \tilde \rho_{n+1}\; d\phi_{n+1} = \tilde \rho_n\; d\phi_n.
\end{equation}
We can now use $(A.1)$ and $(A.3)$ to solve for the correction factor $\alpha$,
through which we can measure exact observables:
\begin{equation}
  \alpha_{n+1} = \alpha_n \frac{d\phi_{n+1}}{ d\phi_n}
                                \frac{\rho_{n+1}}{ \rho_n}
	       = \alpha_n \; {\rm det}
             \left(\partder{M}{\phi}\right)\;\frac{\rho_{n+1}}{ \rho_n},
\end{equation}
where det$(\partial M/\partial\phi)$ is just the Jacobian of the map $\phi_n
\rightarrow \phi_{n+1}$. Thus, by choosing $\alpha_0=1$ at the initial point of
the trajectory, we can compute $\alpha_n$ at all subsequent points along the
trajectory.

Note that $\alpha$ should play a role very similar to the acceptance rate in
HMC.  When $dt$ is small, the distribution $\tilde \rho$ is very close to
$\rho$, so $\alpha \approx 1$ and all of the configurations will be of
approximately equal statistical weight.  This corresponds to high acceptance
rates in HMC.  When $dt$ is large, on the other hand, $\alpha$ will be large
in some regions and small in others, meaning that the simulation spends
significant amounts of time in regions of low statistical importance.  This
is similar to the rejection of many proposed configurations when the
acceptance is low.  Also note that, although in general it is quite difficult,
for our leapfrog algorithm the calculation of det$(\partial M/\partial\phi)$
is straight forward and
only ${\cal O}(volume)$. The correction factor can also be written as
\begin{eqnarray}
    \alpha_n &=& \alpha_o\exp\left\{ \sum_{i=0}^{n-1} {\ell
    n}\left[\partder{q_{i+1}}{q_i}\partder{p_{i+1}}{p_i}\right] - \beta H_i +
   \beta H_0\right\}\\
    &=& \alpha_o\exp\left\{-\beta\left[ {\cal E}_n - {\cal E}_o + O(\Delta
    t^2)\right]\right\},\nonumber
\end{eqnarray}
where ${\cal E}_n$ is the pseudo-energy ~(\ref{eq:pse}) evaluated at $\phi_n$.

We have implemented this algorithm for small harmonic and anharmonic oscillator
systems.  In all runs, we found that, along the trajectory, the magnitude of
$\alpha$ would occasionally rapidly decrease several orders of magnitude and
never recover.  We believe that this is due to an instability in the equations
of motion: if the demons become too large at fixed $\kappa$ and $dt$, they
begin to grow in an unbounded manner.  It is quite likely that this means that
a non-zero $\tilde \rho$ does not exist. We have tried decreasing $dt$ and
various improvements in the equations of motion, but all of these fixes only
decrease the frequency of drops - they do not eliminate them.

In the scheme described above, it is not obvious how to perform a momentum
refresh.  This is because the absolute magnitude of $\alpha$ is not known,
only its ratio to the previous alpha along the trajectory.  We will now show
(again, assuming the existence of $\tilde\rho$) that the correct procedure
is to reset $\alpha$ to $1$ after a gaussian momentum refresh.
Assume, given $\tilde\rho$ and a refreshing scheme, that the probability
of a trajectory beginning at $\phi$ is $P(\phi)$.  In addition, let
$\bar\alpha(\phi)$ be the correctly normalized function $\alpha$ discussed
above and $\alpha(\phi | \phi_0) = \bar\alpha(\phi)/\bar\alpha(\phi_0)$ be
the value of alpha obtained using ($A.4$) along the trajectory from $\phi_0$
to $\phi$ (note that $\alpha(\phi_0 | \phi_0) = 1$). Then
\begin{equation}
  \VEV{\cal O}
    = \frac{ \int d\phi_0 P(\phi_0) \int d\phi {\cal O}\tilde\rho(\phi)
						   \alpha(\phi | \phi_0)}
    { \int d\phi_0 P(\phi_0) \int d\phi         \tilde\rho(\phi)
						   \alpha(\phi | \phi_0)}
   = \frac{N\VEV{ {\cal O}\bar\alpha}'}{ N\VEV{\bar\alpha}'}
\end{equation}
where $N = \int d\phi_0(P(\phi_0)/\bar\alpha(\phi_0))$.  We attempted to test
this scheme on the systems discussed above, but we found that the statistical
errors in the exact demon simulation tended to be larger than the systematic
errors in a similar inexact simulation.  Thus,  it is presently more efficient
to eliminate $dt$ errors by extrapolation techniques than by using the exact
algorithm.

\newpage
\mbox{}

\vspace{2mm}

\noindent{TABLE 1. Comparison of observables between algorithms. For each
$\beta$, the four rows correspond to global demons, HMC-1, HMC-S, HMC-L,
respectively. Measurements were performed on a $64^2$ lattice, with 160K/T
statistics.}

\vspace{4mm}

\begin{tabular}{l@{\extracolsep{2mm}}llllll}\hline
\rule[0cm]{0cm}{4mm}
$\beta$ & $E$ & $C_v$ & $Q\times 100$ & $\chi_Q\times 100$ & $\mid M\mid$ &
                                                     $\chi_M$ \\ [2mm] \hline
\\
0.70    & 0.8287(1)& 0.760(3)&  6.241(2) &  4.49(2)& 0.0484(2)& 12.2(1)\\
        & 0.8299(1)& 0.765(5)&  6.235(2) &  4.50(3)& 0.0487(3)& 12.4(2)\\
        & 0.8297(2)& 0.750(7)&  6.233(2) &  4.40(3)& 0.0493(3)& 12.7(1)\\
        & 0.8300(3)&0.686(14)&  6.231(5) &  4.36(7)& 0.0490(3)& 12.5(1)\\
\\ \hline
\\
0.78    & 0.9573(1)& 0.992(4)&  4.725(2) &  3.92(2)& 0.0683(3)& 24.3(2)\\
        & 0.9577(2)& 0.994(8)&  4.730(3) &  3.93(3)& 0.0683(6)& 24.3(4)\\
        & 0.9573(3)&1.010(12)&  4.731(3) &  3.94(4)& 0.0674(4)& 23.6(3)\\
        & 0.9582(5)&1.061(30)&  4.722(6) &  4.04(9)& 0.0683(5)& 24.3(3)\\
\\ \hline
\\
0.82    & 1.0243(1)& 1.138(5)&  3.981(2) &  3.62(2)& 0.0848(5)& 37.2(4)\\
        & 1.0238(2)& 1.13(1) &  3.995(3) &  3.61(3)& 0.0828(9)& 35.8(8)\\
        & 1.0244(3)& 1.12(2) &  3.987(3) &  3.61(4)& 0.0828(7)& 35.7(6)\\
        & 1.0234(6)& 1.09(4) &  3.996(7) & 3.53(10)& 0.0837(7)& 36.4(6)\\
\\ \hline
\\
1/1.1   & 1.1757(2)& 1.396(7)&  2.431(2) &  2.73(2)& 0.162(1)& 133(2)\\
        & 1.1750(3)& 1.40(2) &  2.448(4) &  2.76(4)& 0.151(3)& 116(4)\\
        & 1.1756(5)& 1.44(4) &  2.443(5) &  2.80(5)& 0.161(2)& 132(3)\\
        &1.1777(10)& 1.38(6) &  2.418(10)& 2.67(11)& 0.166(2)& 140(3)\\
\\ \hline
\\
1/1.04  & 1.2632(3)& 1.512(9)&  1.638(3) &  2.14(2)& 0.295(3)& 409(7)\\
        & 1.2618(5)& 1.50(3) &  1.659(5) &  2.14(4)& 0.276(7)& 370(20)\\
        & 1.2618(7)& 1.50(6) &  1.657(6) &  2.14(6)& 0.281(4)& 374(10)\\
        &1.2630(13)& 1.76(17)&  1.647(12)& 2.33(15)& 0.291(5)& 401(11)\\
\\ \hline
\end{tabular}
\newpage
\mbox{}

\vspace{2mm}

\noindent TABLE 2. Comparison of estimated critical exponents $z$ and prefactor
$A$ for global demons, HMC-1, HMC-S and HMC-L, where $\tau = A \xi^z$.
Measurements were performed on a $64^2$ lattice, with 160K/T statistics.

\vspace{4mm}

\begin{center}
\begin{tabular}{l@{\extracolsep{2cm}}ll}\hline
\rule[0cm]{0cm}{3mm}
       &  $z$   &  $A$ \\ [2mm]\hline
\\
\multicolumn{3}{l}{\em Lattice Magnetization M :}\\
\\
 Global Demons &   1.3& 2.4\\
 HMC-1  &   2.0       & 5.7\\
 HMC-S  &   1.3       & 6.9\\
 HMC-L  &   1.05      & 5.8\\
\\ \hline
\\
\multicolumn{3}{l}{\em Single Spin S :}\\
\\
 Global Demons &   0.8& 1.0\\
 HMC-1  &   1.5       & 1.6\\
 HMC-S  &   1.0       & 1.6\\
 HMC-L  &   1.0       & 4.3\\
\\ \hline
\\
\multicolumn{3}{l}{\em Potential Energy E :}\\
\\
 Global Demons &   0.5& 0.8\\
 HMC-1  &   1.0       & 1.9\\
 HMC-S  &   1.4       & 2.1\\
 HMC-L  &   1.3       & 14 \\
\\ \hline
\\
\multicolumn{3}{l}{\em Topological Charge Q :}\\
\\
 Global Demons &   0.8& 0.8\\
 HMC-1  &   1.0       & 1.4\\
 HMC-S  &   1.3       & 1.6\\
 HMC-L  &   1.3       & 8.2\\
\\ \hline
\end{tabular}
\end{center}
\newpage
\mbox{}

\vspace{2mm}

\noindent{TABLE 3. Auto-correlation times for total magnetization $M$, a single
spin $S$, topological charge $Q$ and internal energy $E$, and the magnetization
correlation length $\xi$. Measurements were performed on a $64^2$ lattice, with
160K/T statistics.}

\vspace{4mm}

\begin{center}
\begin{tabular}{l@{\extracolsep{1.2cm}}ccccc}\hline
\rule[0cm]{0cm}{3mm}
$\beta$ &  $\tau_M$ &  $\tau_S$ &  $\tau_E$ &  $\tau_Q$ & $\xi$ \\ [2mm] \hline
\\
\multicolumn{6}{l}{\em Global Demons~:}\\
\\
0.70    &     6.5   &  1.86  &  1.14  &  1.53 &2.2  \\
0.78    &    11.8   &  2.6   &  1.38  &  1.97 &3.4  \\
0.82    &    16.8   &  3.2   &  1.57  &  2.5  &4.3  \\
1/1.1   &    47.    &  5.9   &  2.2   &  4.5  &9.4  \\
1/1.07  &    62.    &  7.8   &  2.4   &  5.5  &12.8 \\
1/1.04  &   106.    & 10.3   &  3.1   &  8.8  &18.4 \\
1.0     &   124.    & 22.    &  2.8   &  9.5  &31.  \\
\\ \hline
\\
\multicolumn{6}{l}{\em HMC-1~:}\\
\\
0.70    &    27.    &  5.1   &  4.1   &  3.2 &2.2  \\
0.78    &    62.    &  9.2   &  5.9   &  4.8 &3.3  \\
0.82    &   106.    & 12.1   &  7.9   &  6.2 &4.3  \\
1/1.1   &   380     & 36.    & 15.7   & 13.5 &8.9  \\
1/1.04  &  1620     & 99.    & 30.    & 26.  &17.1 \\
\\ \hline
\\
\multicolumn{6}{l}{\em HMC-S~:}\\
\\
0.70    &    17.9  &  3.7  &  6.2  &  4.0  &  2.2  \\
0.78    &    32.   &  5.5  & 12.6  &  8.0  &  3.3  \\
0.82    &    44.   &  7.4  & 18.   & 10.5  &  4.3  \\
1/1.1   &   126.   & 14.   & 50.   & 31.   &  9.3  \\
1/1.04  &   240    & 32.   &132.   & 59.   & 17.5  \\
\\ \hline
\\
\multicolumn{6}{l}{\em HMC-L~:}\\
\\
0.70    &    12.7  &   9.3  &  36.   &  22.   &   2.2  \\
0.78    &    21.   &  15.5  &  66.   &  38.   &   3.4  \\
0.82    &    29.   &  20.   & 136.   &  61.   &   4.3  \\
1/1.1   &    64.   &  40.   & 144.   & 132.   &   9.5  \\
1/1.04  &   114.   &  84.   & 760    & 350    &  17.7  \\
\\ \hline
\end{tabular}
\end{center}
\newpage
\begin{center}
 {\large FIGURES }
\end{center}

\begin{enumerate}

\item  {Time evolution of $10^3$ connected points evolving
according to `canonical' rather than microcanonical dynamics for (a) the 1-d
harmonic oscillator $H=(p^2+q^2)/2$ (left column; $t=0$ points lie on the
unit circle), and  (b) the  $SU(2)$ Hamiltonian $H=J_z^2/2$ (right column;
$t=0$ points lie on the circle $J_z=0.5$), both at $\beta=1$. The phase  space
in (a) is $(q,p)$ and in (b) it is the sphere parameterized by
$(J_x,J_y,J_z)$. The rapid spreading of neighboring points is characteristic of
global demon dynamics. The characteristic time scale is $t\sim 1$.}
\item {Finite time step extrapolations, demonstrating the global
$O(dt^2)$ leapfrog error for (a) potential energy, (b) topological charge and
(c) magnetization.}
\item {Diffusiveness of a global demon trajectory versus the
coupling  strength, for selected values of $\beta$. $\Delta S$ is measured at
intervals of $T=1$. The `optimal' value of $\difu=\sigma$ is indicated by the
dashed line. As can be seen, the dynamics is not strongly affected by critical
slowing down. Saturation occurs when $\kappa_\alpha\sim O(\sqrt{2n}) \sim 23$.
The optimal coupling can be seen to be independent of both $\beta$ and the
phase transition.}
\item {Diffusiveness of the global demons versus $\beta$. The dip
at $\beta\sim 1$ is a result of the Kosterlitz-Thouless phase transition and
critical slowing down. The dynamics is not strongly effected by the transition,
so no additional tuning is required, and $\kappa$ can be taken as fixed for all
$\beta$.}
\item {Measurement dependence on the coupling $\kappa$ at
$\beta=1/1.1$.}
\item {Diffusiveness of a global demon trajectory a function of
the trajectory length $T$, at $\beta = 1$, $\kappa_1=\kappa_2=16$. Saturation
can be seen, indicating an optimal trajectory length of $T\sim O(1)$.}
\item {Autocorrelation functions  for potential energy,
topological charge, magnetization and spin, at $\beta=1/1.1$ with
$\kappa_1=\kappa_2=1$ (dots) and $\kappa_1=\kappa_2=16$ (solid). As can be
seen, the ringing vanishes as the coupling increases.}
\item {Autocorrelation functions for potential energy,
topological charge, magnetization and spin, at $\beta=1/1.1$ for global demons
with $\kappa_1=\kappa_2=16$ (solid), HMC-1 (dots), HMC-S (crosses) and HMC-L
(boxes). For global demons and HMC-1, measurements are made every $t=1$, while
for HMC-S,L, the boxes and crosses  indicate the actual number of data points.}
\item {Lattice magnetization auto-correlation function at
selected temperatures for global demons (solid), HMC-1 (dots), HMC-S (dashes)
and HMC-L (boxes). The time axis for $\beta=0.7$ has been scaled by a factor of
0.1 to magnify the short time behavior.}
\item {The behavior of $\tau=a\xi^z$ is plotted near the phase
transition on a $64^2$ lattice for (a) total magnetization, (b) the spin at the
origin, (c) topological charge and (d) potential energy. In each figure, we
indicate the results for HMC (boxes), HMC$-s$ (diamonds), HMC$-l$ (crosses) and
global demons (circles). Critical exponents can be extracted from a linear fit.
The results of the numerical fits are given in Table 2. Results are on a $64^2$
lattice with $160K/T$ statistics.}
\item {Decorrelation times versus coupling strength for $\beta =
0.5$ (squares), $0.7$ (diagonal crosses), $1/1.1$ (diamonds), $2.0$ (vertical
crosses) and $4.0$ (circles) for (a) total magnetization, (b) a single spin,
(c) topological charge and (d) potential energy.}
\item {Decorrelation times versus $\beta$ for couplings
$\kappa_1=\kappa_2=1$ (crosses), $4$ (diamonds), $16$ (squares) and $64$
(circles) for (a) total magnetization, (b) a single spin, (c) topological
charge and (d) potential energy. The rise at $\beta\sim 1$ is critical slowing
down, and is especially evident for the energy and topological charge.}

\end{enumerate}


\begin{thebibliography}{20}
\bibitem{orbit}The identification of stable periodic orbits is less clear. In
         practice such situations have not been encountered. In any case, this
         can be repaired by a periodic momentum refresh.
\end{thebibliography}
\end{document}